\documentstyle[aps,pra,multicol,amssymb]{revtex}
\begin{document}
\draft
\title{{Two-particle entanglement as a property of three-particle entangled states}\thanks{This paper was originally published in: J. L. Cereceda, Phys. Rev. A {\bf 56}, 1733-1738 (1997).}}
\author{Jos\'{e} L. Cereceda}
\address{C/Alto del Le\'{o}n 8, 4A, 28038 Madrid, Spain}
\date{27 May 1999}
\maketitle
\begin{abstract}
In a recent article [Phys. Rev. A {\bf 54}, 1793 (1996)] Krenn and Zeilinger investigated the conditional two-particle correlations for the subensemble of data obtained by selecting the results of the spin measurements by two observers 1 and 2 with respect to the result found in the corresponding measurement by a third observer. In this paper we write out explicitly the condition required in order for the selected results of observers 1 and 2 to violate Bell's inequality for general measurement directions $\vec{e}_1$, $\vec{e}\, ^{\prime}_{1}$, $\vec{e}_2$, $\vec{e}\,^{\prime}_{2}$, and $\vec{e}_3$. It is shown that there are infinitely many sets of directions giving the maximum level of violation. Further, we extend the analysis by the authors to the class of triorthogonal states $|\Psi \rangle = c_1 |z_1 \rangle |z_2 \rangle |z_3 \rangle + c_2 |-z_1 \rangle |-z_2 \rangle |-z_3 \rangle$. It is found that a maximal violation of Bell's inequality occurs provided the corresponding three-particle state yiels a direct (``all or nothing'') nonlocality contradiction.

\end{abstract}

\begin{multicols}{2}
\section{Introduction}

Recently, Krenn and Zeilinger \cite{KZ} (hereafter referred to as KZ) have shown that situations can arise where the property of entanglement of quantum systems is itself an entangled property. For that purpose they considered a three-particle system described by the Greenberger-Horne-Zeilinger (GHZ) state \cite{GHZ}
\begin{equation}
|\Psi \rangle = \frac{1}{\sqrt{2}} (|z+ \rangle_1 |z+ \rangle_2 |z+ \rangle_3
+ |z- \rangle_1 |z- \rangle_2 |z- \rangle_3),
\end{equation}
where $|z+ \rangle_i$ ($|z- \rangle_i$) represents a state of spin-up (-down) for particle $i$, $i=1,2,3$, along some quantization $z$ axis which in general will differ from one particle to the other. We can regard the three particles as flying apart from a common source, each of them subsequently entering its own Stern-Gerlach apparatus oriented along an arbitrary measurement direction $\vec{e}_i$ in three-dimensional space, this direction being specified by the polar and azimuthal angles $\vartheta_i$ and $\varphi_i$. It is assumed that at the time of measurement the particles 1, 2, and 3 may be arbitrarily far apart so that the acts of measurement by respective observers 1, 2, and 3 can be considered to have a spacelike separation. KZ first showed that the correlation function $E_{12}$ obtained by unconditionally averaging the products of the results of the measurements on particles 1 and 2 factorizes into a product of two functions, one of them related to particle 1 and the other related to particle 2, so that we can always think of such results as being clasically correlated. Next the authors examined a situation in which observer 3, independent of observers 1 and 2, performs spin measurements on particle 3 along direction $\vec{e}_3$. The results obtained by observer 3 are then used to classify the results of observers 1 and 2 into two distinct subensembles: whenever a result $+1$ ($-1$) is found for particle 3 in a particular run of the experiment, the corresponding results for particles 1 and 2 are assigned to subensemble $+$ ($-$). KZ demonstrated that an entanglement between particles 1 and 2 indeed occurs by showing that for certain measurement directions $\vec{e}_3$ the resulting correlation function $E^{+}_{12}$ ($E^{-}_{12}$) for subensemble $+$ ($-$) can yield a violation of Bell's inequality. Since the degree of entanglement within either subensemble $+$ or $-$ depends on the measurement direction $\vec{e}_3$, and due to the fact that the spin measurements are carried out on the particles in spacelike separated regions, KZ came to the conclusion that the property of entanglement depends on the whole measurement context and therefore becomes an entangled property itself.

In this paper we state the general condition for the violation of a Bell inequality involving the selected results in either one of the above-defined subensembles. In particular, constraints for a maximal violation are given. This is done in a way that explicitly shows the dependence of  the entanglement of the subensembles on the setting of a measuring apparatus which can be located in a spacelike separated region. Furthermore, we extend the analysis of KZ to include a more general type of three-particle states than that appearing in Eq.\ (1). Specifically, we shall consider the class of states which can be written in the triorthogonal form \cite{Elby-Bub}
\begin{equation}
|\Psi \rangle = c_1 |z_1 \rangle |z_2 \rangle |z_3 \rangle
+ c_2 |-z_1 \rangle |-z_2 \rangle |-z_3 \rangle,
\end{equation}
where, for simplicity, the coefficients $c_1$ and $c_2$ are chosen real, with $c^{2}_{1} + c^{2}_{2} =1$, and where $|z_i \rangle$ ($|-z_i \rangle$) denotes the eigenvector of the spin operator along the $z$ axis for particle $i$, with eigenvalue $z_i =\pm1$ ($-z_i = \mp1$). Clearly, the state in Eq.\ (2) reduces to the GHZ state (1) when we put $c_1 =c_2 = 1/{\sqrt{2}}$, and $z_1 = z_2 = z_3 =+1$. However, for the class of states considered, it is shown that no violations of local realism can arise within the unconditional ensemble constructed by selecting the results obtained by two of the observers (say, by observers 1 and 2) irrespective of the result obtained in the corresponding measurement by the third observer. As will presently be seen, this unrestricted selection of the data provided by the results of observers 1 and 2 involves the incoherent mixture of two pure states $|\Psi^{+}_{12}\rangle$ and $|\Psi^{-}_{12}\rangle$ for particles 1 and 2, with weighting factors given by the probability of getting the result $z_3$ and $-z_3$, respectively, in a spin measurement on particle 3 along the axis $\vec{e}_3 (\vartheta_{3},\varphi_{3})$. Such a mixture, however, turns out to be completely equivalent to a mixture of two product states for particles 1 and 2, and then no violation of a Bell inequality will occur for the above-defined unconditional ensemble.

\section{Unconditional two-particle correlations for the class of triorthogonal states}

In order to determine under which conditions the correlation between the measurement results for particles 1 and 2 can lead to a violation of Bell's inequality it is convenient to introduce another set of basis states $\{|z_3 \rangle^{\textstyle \ast},|-z_3 \rangle^{\textstyle\ast}\}$ for particle 3, related to the original basis vectors by
\begin{mathletters}
\begin{equation}
|z_{3} \rangle = \cos \frac{\vartheta_3}{2} e^{iz_{3}\varphi_{3}/2} |z_{3} \rangle^{\textstyle \ast} - z_{3} \sin \frac{\vartheta_3}{2} e^{iz_{3}\varphi_{3}/2} |-z_{3} \rangle^{\textstyle\ast},  
\end{equation}
\begin{equation}
|-z_{3} \rangle = z_{3} \sin \frac{\vartheta_3}{2} e^{-iz_{3}\varphi_{3}/2} |z_{3} \rangle^{\textstyle\ast}
+ \cos \frac{\vartheta_3}{2} e^{-iz_{3}\varphi_{3}/2} |-z_{3} \rangle^{\textstyle\ast}. 
\end{equation}
\end{mathletters}
Using Eqs.\ (3a) and (3b) we can rewrite the three-particle state (2) as
\begin{equation}
|\Psi \rangle = p^{1/2}_{+}|\Psi^{+}_{12} \rangle |z_{3} \rangle^{\textstyle\ast}
+ p^{1/2}_{-}|\Psi^{-}_{12} \rangle |-z_{3} \rangle^{\textstyle\ast},
\end{equation}
where
\begin{mathletters}
\begin{eqnarray}
|\Psi^{+}_{12} \rangle &=& p^{-1/2}_{+} \left( c_1 \cos \frac{\vartheta_{3}}{2}
|z_1 \rangle |z_2 \rangle \right. \nonumber   \\
&& + \left. c_2 z_3 \sin \frac{\vartheta_{3}}{2} e^{-iz_{3}\varphi_{3}}
|-z_1 \rangle |-z_2 \rangle \right),
\end{eqnarray}
\begin{eqnarray}
|\Psi^{-}_{12} \rangle &=& p^{-1/2}_{-} \left( -c_1 z_3 \sin \frac{\vartheta_{3}}{2}
|z_1 \rangle |z_2 \rangle \right. \nonumber   \\
&& + \left. c_2 \cos \frac{\vartheta_{3}}{2} e^{-iz_{3}\varphi_{3}}
|-z_1 \rangle |-z_2 \rangle \right),
\end{eqnarray}
\end{mathletters}
and
\begin{mathletters}
\begin{eqnarray}
p_+ &=& c^{2}_{1} \cos^2 \left( \frac{\vartheta_{3}}{2} \right)
+ c^{2}_{2} \sin^2 \left( \frac{\vartheta_{3}}{2} \right),  \\
p_- &=& c^{2}_{1} \sin^2 \left( \frac{\vartheta_{3}}{2} \right)
+ c^{2}_{2} \cos^2 \left( \frac{\vartheta_{3}}{2} \right).
\end{eqnarray}
\end{mathletters}
Since the basis vectors $|\pm z_3 \rangle$ represent definite spin $\pm z_3$ along the $z$ axis for particle 3, the basis vectors $|\pm z_3 \rangle^{\textstyle\ast}$ represent definite spin $\pm z_3$ for particle 3 along the direction characterized by the angles $\vartheta_3$ and $\varphi_3$. In view of Eq.\ (4) it thus follows that, provided $\vartheta_3 \neq n\pi$, $n=0,\pm1,\pm2,\ldots ,$ a spin measurement on particle 3 along (an otherwise arbitrary) direction $\vec{e}_3 (\vartheta_{3},\varphi_{3})$ leaves the particles 1 and 2 in an entangled state. Specifically, if the measurement result for particle 3 is found to be $z_3$, the normalized two-particle state for particles 1 and 2 becomes $|\Psi^{+}_{12}\rangle$, whereas for a measurement result equal to $-z_3$ it becomes $|\Psi^{-}_{12}\rangle$. However, from expression (4), we can see that the probability of obtaining the result $z_3$ ($-z_3$) in a spin measurement on particle 3 along the axis $\vec{e}_3 (\vartheta_{3},\varphi_{3})$ is $p_+$ ($p_-$). Of course we have $p_+ + p_- =1$. It will further be noted that the states $|\Psi^{+}_{12}\rangle$ and $|\Psi^{-}_{12}\rangle$ are orthogonal to each other provided that $|c_1|=|c_2|=1/{\sqrt{2}}$.

If we analyze the results of observers 1 and 2 only when observer 3 obtains the result $z_3$ ($-z_3$) (i.e., if we restrict ourselves to either one of subensembles $+$ or $-$) then the adequate state describing the spin of particles 1 and 2 will be the pure state $|\Psi^{+}_{12}\rangle$ ($|\Psi^{-}_{12}\rangle$). However, if we attemp to analyze the results of observers 1 and 2 irrespective of what result happens to be measured by observer 3 in the corresponding measurement along direction $\vec{e}_3$ (i.e., if we consider the total ensemble formed out of subensembles $+$ and $-$), then the appropriate state accounting for particles 1 and 2 will consist of a mixture of both pure states $|\Psi^{+}_{12}\rangle$ and $|\Psi^{-}_{12}\rangle$ with respective weights $p_+$ and $p_-$; that is,
\begin{equation}
\rho_{12} = p_+ |\Psi^{+}_{12}\rangle \langle \Psi^{+}_{12}|
+ p_- |\Psi^{-}_{12}\rangle \langle \Psi^{-}_{12}| .
\end{equation}
It is a rather simple matter to show that the density matrix (7) can also be decomposed in the form
\begin{equation}
\rho_{12} = c^{2}_{1} |\phi^{+}_{12}\rangle \langle \phi^{+}_{12}|
+ c^{2}_{2} |\phi^{-}_{12}\rangle \langle \phi^{-}_{12}| ,
\end{equation}
with $|\phi^{+}_{12}\rangle = |z_1\rangle |z_2\rangle$, and $|\phi^{-}_{12}\rangle = |-z_1\rangle |-z_2\rangle$. It should be noted at this point that both expressions (7) and (8) for $\rho_{12}$ can also be obtained by taking the partial trace of the density operator $|\Psi\rangle \langle\Psi|$ over the states corresponding to particle 3. So, on choosing the basis states for particle 3 to be $\{|z_3 \rangle^{\textstyle\ast},|-z_3 \rangle^{\textstyle\ast}\}$, we can put
\[
\rho_{12} = { }^{\textstyle\ast} \langle z_3 |\Psi\rangle\langle\Psi|z_3 \rangle^{\textstyle\ast}
+ { }^{\textstyle\ast} \langle -z_3 |\Psi\rangle\langle\Psi|-z_3 \rangle^{\textstyle\ast} .  
\]
Replacing now $|\Psi\rangle$ by the state in Eq.\ (4) we obtain quickly the density matrix (7). Likewise, by tracing over the states $\{|z_3 \rangle,|-z_3 \rangle\}$ we get
\[
\rho_{12} = \langle z_3 |\Psi\rangle\langle\Psi|z_3 \rangle
+ \langle -z_3 |\Psi\rangle\langle\Psi|-z_3 \rangle ,  
\]
which can be identified with the density matrix (8) upon substitution of $|\Psi\rangle$ by the state vector (2).

The important point about the decomposition in Eq.\ (8) is that, as both $|\phi^{+}_{12}\rangle$ and $|\phi^{-}_{12}\rangle$ are product states, none of them violates Bell's inequalities, and then the same will be true for the mixed state $\rho_{12}$ \cite{Barnett-Phoenix}. Indeed, one can easily find that the quantum prediction for the unconditional correlation function of the results of spin measurements on particles 1 and 2 along directions $\vec{e}_1$ and $\vec{e}_2$, respectively, is
\begin{eqnarray}
E_{12}(\vec{e}_1,\vec{e}_2) &=& \text{Tr}[\rho_{12} \sigma(\vec{e}_1)\otimes
\sigma(\vec{e}_2)]  \nonumber  \\
&=& \pm \cos \vartheta_1 \cos \vartheta_2 ,
\end{eqnarray}
where the matrix $\rho_{12}$ used in Eq.\ (9) stands for either one of expressions (7) or (8), and where the $+$ ($-$) sign applies for $\text{sgn}z_1 = \text{sgn}z_2$ ($\text{sgn}z_1 \neq \text{sgn}z_2$). So, in order to search for genuinely quantal correlations involving the pair of particles 1 and 2 it is necessary to consider the measurement results pertaining to either subensemble $+$ or $-$ separately. As a result, in the following we shall deal with the pure state $|\Psi^{+}_{12}\rangle$ or $|\Psi^{-}_{12}\rangle$, rather than with the mixed state $\rho_{12}$.

It is worth noting that the inability of the entangled three-particle state (2) to yield nonlocal correlations involving the unconditional measurement results for two of the particles can be traced back to the orthogonality of the states $|z_i\rangle$ and $|-z_i\rangle$, for each $i=1,2,$ and 3. So we can say that the lack of such nonlocal, unconditional two-particle correlations indeed constitutes a significant feature inherent in the class of triorthogonal states. In Sec.\ IV we shall generalize this result to the class of $n$-orthogonal states.

\section{Maximal violation of the CHSH inequality: necessary conditions}

The Clauser-Horne-Shimony-Holt (CHSH) form of the Bell inequality is \cite{CHSH}
\begin{equation}
|\langle B_{\text{CHSH}}\rangle | \leq 2 \, ,
\end{equation}
where $\langle B_{\text{CHSH}}\rangle$ denotes the expectation value of the Bell operator \cite{BMR}
\begin{eqnarray}
B_{\text{CHSH}} &\equiv & \sigma(\vec{e}_1) \otimes [\sigma(\vec{e}_2) + \sigma(\vec{e}\,%
^{\prime}_{2})] + \sigma(\vec{e}\,^{\prime}_{1}) \nonumber  \\
&& \otimes \, [\sigma(\vec{e}_2) - \sigma(\vec{e}\,^{\prime}_{2})] \, ,
\end{eqnarray}
with $\vec{e}_i$ and $\vec{e}\,^{\prime}_{i}$ denoting two different alternative directions for spin measurements on particle $i$. As long as $\vartheta_3 \neq n\pi$, either two-particle state (5a) or (5b) will be able to violate the CHSH inequality for a suitable choice of measurement directions $\vec{e}_1$, $\vec{e}\,^{\prime}_{1}$, $\vec{e}_2$, and $\vec{e}\,^{\prime}_{2}$ \cite{Gisin,PR}. Now, the quantum prediction for the correlation function $E^{+}_{12}$ ($E^{-}_{12}$) associated with subensemble $+$ ($-$) is given by
\begin{eqnarray}
E^{\pm}_{12}(\vec{e}_1,\vec{e}_2) &=& \langle \Psi^{\pm}_{12}|\sigma(\vec{e}_1) \otimes \sigma(\vec{e}_2) |\Psi^{\pm}_{12}\rangle  \nonumber \\
&=& \gamma \cos\vartheta_1 \cos\vartheta_2 \pm z_3 (c_1 c_2 /p_{\pm}) \sin\vartheta_1
\sin\vartheta_2  \nonumber  \\
&& \times\sin\vartheta_3 \cos (\varphi_1 +\gamma\varphi_2 +z_1 z_3 \varphi_3),
\end{eqnarray}
where $\gamma$ is a sign factor with value $+1$ ($-1$) for $\text{sgn}z_1 = \text{sgn}z_2$ ($\text{sgn}z_1 \neq \text{sgn}z_2$). Incidentally, it will be noted that for $\vartheta_3 = n\pi$ we have $E^{\pm}_{12} = \gamma \cos\vartheta_1 \cos\vartheta_2$, and hence it is not possible a violation of Bell's inequality for any choice of directions $\vec{e}_1$, $\vec{e}\,^{\prime}_{1}$, $\vec{e}_2$, $\vec{e}\,^{\prime}_{2}$. Of course this arises from the fact that the superposition state (2) reduces to the unentangled term $|z_1 \rangle |z_2 \rangle |z_3 \rangle$ or $|-z_1 \rangle |-z_2 \rangle |-z_3 \rangle$ whenever a spin measurement along the $z$ axis is performed on particle 3. Further, as expected, for either $c_1 =0$ or $c_2 =0$ (i.e., for product states) no violation can happen. Therefore, for general directions $\vec{e}_1$, $\vec{e}\,^{\prime}_{1}$, $\vec{e}_2$, $\vec{e}\,^{\prime}_{2}$, and $\vec{e}_3$, the relevant predictions by quantum mechanics will violate inequality (10) if
\begin{eqnarray}
|\gamma &[& \cos\vartheta_1 \cos\vartheta_2 + \cos\vartheta_1 
\cos\vartheta^{\prime}_{2} + \cos\vartheta^{\prime}_{1} \cos\vartheta_2  \nonumber \\
&& - \cos\vartheta^{\prime}_{1} \cos\vartheta^{\prime}_{2}]
\pm z_3 (c_1 c_2 /p_{\pm}) [\sin\vartheta_1 \sin\vartheta_2 \sin\vartheta_3 \nonumber   \\
&& \times \cos (\varphi_1 +\gamma \varphi_2+z_1 z_3 \varphi_3) +\sin\vartheta_1 \sin\vartheta^{\prime}_{2} \sin\vartheta_3 \nonumber  \\
&& \times \cos (\varphi_1 +\gamma\varphi^{\prime}_{2} +z_1 z_3 \varphi_3) 
+ \sin\vartheta^{\prime}_{1} \sin\vartheta_2 \sin\vartheta_3 \nonumber  \\
&& \times \cos (\varphi^{\prime}_{1} +\gamma\varphi_2 +z_1 z_3 \varphi_3)  
- \sin\vartheta^{\prime}_{1} \sin\vartheta^{\prime}_{2} \sin\vartheta_3 \nonumber  \\
&& \times \cos (\varphi^{\prime}_{1} +\gamma\varphi^{\prime}_{2} +z_1 z_3 \varphi_3)] | > 2.
\end{eqnarray}
It should be realized that, by suitable choosing the $z$ axis for each of the particles, any entangled state for two spin-$\frac{1}{2}$ particles can be put into the form displayed by either one of Eqs.\ (5a) or (5b). So the above condition (13) for the violation of the CHSH inequality turns out to be completely general as far as pure states are concerned. For the special case in which $\vartheta^{\prime}_{1} = \vartheta_1$, $\vartheta^{\prime}_{2} = \vartheta_2$, $\varphi^{\prime}_{1} = \varphi_1 +\pi/2$, $\gamma\varphi^{\prime}_{2} = \gamma\varphi_2 +\pi/2$, and $\varphi_1 +\gamma\varphi_2 +z_1 z_3 \varphi_3 = 3\pi/4 +n\pi$, condition (13) simplifies to
\begin{equation}
|\gamma \cos\vartheta_1 \cos\vartheta_2 \pm \mu z_3 (c_1 c_2 /p_{\pm}) \sqrt{2}
\sin\vartheta_1 \sin\vartheta_2 \sin\vartheta_3 | >1,
\end{equation}
where $\mu$ is a sign factor equal to $+1$ ($-1$) for $n$ odd (even). From expression (14) it follows that, as long as $\vartheta^{\prime}_{1} = \vartheta_1$ and $\vartheta^{\prime}_{2} = \vartheta_2$, a maximal violation of the CHSH inequality occurs provided that (i) $|c_1|=|c_2|=1/\sqrt{2}$, and (ii) $\vartheta_1 =\vartheta_2 =\vartheta_3 =\pi/2$. Clearly, condition (ii) entails that all measurement directions $\vec{e}_1$, $\vec{e}\,^{\prime}_{1}$, $\vec{e}_2$, $\vec{e}\,^{\prime}_{2}$, and $\vec{e}_3$ must lie in the $x$-$y$ plane.

A few remarks should be added here. In the first place, it is to be noted that the requirements 
$|c_1|=|c_2|=1/\sqrt{2}$ and $\vartheta_3 =\pi/2$ together imply that either state (5a) or (5b) is maximally entangled. This is consistent with the fact according to which a maximally entangled state not only gives the maximum violation of the CHSH inequality but also gives the {\it largest\/} violation attainable for any pairs of four spin observables $\sigma(\vec{e}_1)$, $\sigma(\vec{e}\,^{\prime}_{1})$, $\sigma(\vec{e}_2)$, and $\sigma(\vec{e}\,^{\prime}_{2})$ (provided $\vec{e}_1 \!\nparallel \! \vec{e}\,^{\prime}_{1}$, and $\vec{e}_2 \!\nparallel \! \vec{e}\,^{\prime}_{2}$) \cite{Kar}. Therefore, both condition (i) and the requirement $\vartheta_3 = \pi/2$ turn out to be absolutely necessary in order to achieve the largest violation, no matter what the orientation of $\vec{e}_1$, $\vec{e}\,^{\prime}_{1}$, $\vec{e}_2$, and $\vec{e}\,^{\prime}_{2}$ may be. However, for the special case leading to Eq.\ (14) we have that the two measurement directions $\vec{e}_i$ and $\vec{e}\,^{\prime}_{i}$ for particle $i$ ($i=1,2$) giving the maximal violation are perpendicular between themselves. That this orthogonality condition is not incidental can be seen by computing the square of the Bell operator (11). This is given by \cite{Kar}
\begin{equation}
B^{2}_{\text{CHSH}} = 4(I+\sin\theta_1 \sin\theta_2 \, \sigma_{\perp 1} \otimes
\sigma_{\perp 2}),
\end{equation}
where $\theta_i$ is the angle included between the vectors $\vec{e}_i$ and $\vec{e}\,^{\prime}_{i}$, and $\sigma_{\perp i}$ denotes the spin operator for particle $i$ along the direction perpendicular to the plane containing $\vec{e}_i$ and $\vec{e}\,^{\prime}_{i}$. From Eq.\ (15) it follows that the largest eigenvalue of $B_{\text{CHSH}}$ is (in terms of the absolute value) $\lambda_l = 2(1+|\sin\theta_1 \sin\theta_2|)^{1/2}$, so that a maximum violation is obtained when both $\theta_1$ and $\theta_2$ are $\pi/2$ (mod $\pi$). The eigenvector associated with $\lambda_l$ will consist of a superposition of the states $|\sigma_{\perp 1}\rangle|\sigma_{\perp 2}\rangle$ and $|-\sigma_{\perp 1}\rangle|-\sigma_{\perp 2}\rangle$ with an appropriate relative phase between them, where $|\sigma_{\perp i}\rangle$ ($|-\sigma_{\perp i}\rangle$) denotes the eigenvector of $\sigma_{\perp i}$ with eigenvalue $\sigma_{\perp i}=\pm 1$ ($-\sigma_{\perp i}=\mp 1$), and where $\text{sgn}\sigma_{\perp 1} =\text{sgn}\sigma_{\perp 2}$ ($\text{sgn}\sigma_{\perp 1} \neq \text{sgn}\sigma_{\perp 2}$) for $\text{sgn}(\sin\theta_1) = \text{sgn}(\sin\theta_2)$ [$\text{sgn}(\sin\theta_1) \neq \text{sgn}(\sin\theta_2)$]. As was mentioned, such a superposition state has to be completely entangled in order to get the largest possible violation \cite{Kar}.

A question that naturally arises is whether the above conditions  on the measurement directions, namely, $\vartheta_1 = \vartheta^{\prime}_{1} =\pi/2$, $\vartheta_2 = \vartheta^{\prime}_{2} =\pi/2$, $\varphi^{\prime}_{1} = \varphi_1 +\pi/2$, $\gamma\varphi^{\prime}_{2} = \gamma\varphi_2 +\pi/2$, and $\varphi_1 +\gamma\varphi_2 +z_1 z_3 \varphi_3 = 3\pi/4 +n\pi$ exhaust all the possibilities for a maximal violation of the CHSH inequality. The answer is certainly no. Indeed, there are infinitely many ways of achieving such a maximum. This follows directly from the fact that the Schmidt decomposition of a maximally entangled state is not unique. So, for example, consider the state (5a) with $c_1 = -c_2 = 1/\sqrt{2}$, $\vartheta_3 =\pi/2$, $\varphi_3 =0$, $z_1 =-z_2 =+1$, and $z_3 =+1$. The resulting singlet state can be equally expressed in an infinite number of alternative forms by replacing the quantization $z$ axis with any other unit vector ${\bf{\text n}}$ in three-dimensional space. Put it, for instance, as
\begin{equation}
|\Psi^{+}_{12}\rangle = \frac{1}{\sqrt{2}} (|y+\rangle_1  |y-\rangle_2
- |y-\rangle_1 |y+\rangle_2 ),
\end{equation}
where $|y+\rangle_1$ represents spin-up along the $y$ axis for particle 1. Hence, according to the previous paragraph, there must be measurement directions $\vec{e}_1$, $\vec{e}\,^{\prime}_{1}$, $\vec{e}_2$, and $\vec{e}\,^{\prime}_{2}$ in the $x$-$z$ plane (with $\vec{e}_1 \perp \vec{e}\,^{\prime}_{1}$ and $\vec{e}_2 \perp \vec{e}\,^{\prime}_{2}$) such that inequality (10) is maximally violated for the singlet state. As an example, take the choice $\vartheta_1 =0$, $\varphi_1 =0$, $\vartheta^{\prime}_{1}=\pi/2$, $\varphi^{\prime}_{1}=0$, $\vartheta_2 =\pi/4$, $\varphi_2 =0$, $\vartheta^{\prime}_{2}=-\pi/4$, and $\varphi^{\prime}_{2}=0$. Substituting these values (together with $c_1 = -c_2 = 1/\sqrt{2}$, $\vartheta_3 =\pi/2$, $\varphi_3 =0$, $\gamma =-1$, and $z_3 =+1$) into the left-hand side of inequality (13) gives the maximum violation $2\sqrt{2}$. Since the unit vector ${\bf{\text n}}$ is quite arbitrary, we conclude that there are infinitely many sets of directions $\vec{e}_1$, $\vec{e}\,^{\prime}_{1}$, $\vec{e}_2$, and $\vec{e}\,^{\prime}_{2}$ satisfying the equality
\begin{eqnarray}
|&& \cos \vartheta_1 \cos\vartheta_2 + \cos\vartheta_1 
\cos\vartheta^{\prime}_{2} + \cos\vartheta^{\prime}_{1} \cos\vartheta_2  \nonumber  \\
&& - \cos\vartheta^{\prime}_{1} \cos\vartheta^{\prime}_{2}
+ \sin\vartheta_1 \sin\vartheta_2 \cos(\varphi_1 - \varphi_2) \nonumber  \\
&& + \sin\vartheta_1 \sin\vartheta^{\prime}_{2} \cos(\varphi_1 - \varphi^{\prime}_{2})
+ \sin\vartheta^{\prime}_{1} \sin\vartheta_2  \nonumber  \\
&& \times \cos(\varphi^{\prime}_{1} - \varphi_2) - \sin\vartheta^{\prime}_{1}
\sin\vartheta^{\prime}_{2} \cos(\varphi^{\prime}_{1} - \varphi^{\prime}_{2})|
= 2 \sqrt{2}.
\end{eqnarray}
Another example which provides the maximum level of violation is $\varphi_1 =\varphi^{\prime}_{1} =\varphi_2 =\varphi^{\prime}_{2} =\varphi_{0}$, $\vartheta_1 =\vartheta_{0} - \pi/4$, $\vartheta^{\prime}_{1}=\vartheta_0 + \pi/4$, $\vartheta_2 =\vartheta_0$, and $\vartheta^{\prime}_{2}=\vartheta_0 -\pi/2$, with $\varphi_0$ and $\vartheta_0$ taking on any arbitrary value. Note that this example includes the previous one when we make $\varphi_0 =0$ and $\vartheta_0 =\pi/4$. In any case, as was shown, to achieve the maximum violation it is necessary that the vectors $\vec{e}_i$ and $\vec{e}\,^{\prime}_{i}$ ($i=1,2$) be perpendicular between themselves. This orthogonality condition, however, is not a sufficient one, as is clear from the preceding example (indeed, for this example, in addition to this condition it is necessary that $\vartheta_2 =\vartheta_1 +\pi/4$). Note also that the directions giving the largest violation of the CHSH inequality for the singlet state must satisfy in all cases the constraint $\text{sgn}(\sin\theta_1) \neq \text{sgn}(\sin\theta_2)$. This is because the singlet state is an eigenvector of the operator $\sigma_{n1}\otimes\sigma_{n2}$ (where $\sigma_{ni}$ denotes the spin operator along the ${\bf{\text n}}$ axis for particle $i$) with eigenvalue $-1$, and so, from Eq.\ (15), the factors $\sin\theta_1$ and $\sin\theta_2$ must be opposite in sign in order to get the largest violation.

Consider now the state (5a) with $c_1 = c_2 = 1/\sqrt{2}$, $\vartheta_3 =\pi/2$, $\varphi_3 =0$, $z_1 =-z_2 =+1$, and $z_3 =+1$. Unlike the singlet state, the resulting triplet state is not rotationally invariant. However, it can also be written in an infinity of equivalent biorthogonal forms. Put it, for instance, as
\begin{equation}
|\Psi^{+}_{12}\rangle = \frac{1}{\sqrt{2}} (|x+\rangle_1  |x+\rangle_2
- |x-\rangle_1 |x-\rangle_2 ),
\end{equation}
where $|x+\rangle_1$ represents spin-up along the $x$ axis for particle 1. So, there will be measurement directions $\vec{e}_1$, $\vec{e}\,^{\prime}_{1}$, $\vec{e}_2$, and $\vec{e}\,^{\prime}_{2}$ (with $\vec{e}_1 \perp \vec{e}\,^{\prime}_{1}$ and $\vec{e}_2 \perp \vec{e}\,^{\prime}_{2}$) in the $y$-$z$ plane allowing maximal violation of inequality (10) for the triplet state. As an example, take the choice $\vartheta_1 =0$, $\varphi_1 =\pi/2$, $\vartheta^{\prime}_{1}=-\pi/2$, $\varphi^{\prime}_{1}=\pi/2$, $\vartheta_2 =\pi/4$, $\varphi_2 =\pi/2$, $\vartheta^{\prime}_{2}=-\pi/4$, and $\varphi^{\prime}_{2}=\pi/2$. As may easily be checked, these values fulfill the equality
\begin{eqnarray}
|&& \cos \vartheta_1 \cos\vartheta_2 + \cos\vartheta_1 
\cos\vartheta^{\prime}_{2} + \cos\vartheta^{\prime}_{1} \cos\vartheta_2  \nonumber  \\
&& - \cos\vartheta^{\prime}_{1} \cos\vartheta^{\prime}_{2}
- \sin\vartheta_1 \sin\vartheta_2 \cos(\varphi_1 - \varphi_2) \nonumber  \\
&& - \sin\vartheta_1 \sin\vartheta^{\prime}_{2} \cos(\varphi_1 - \varphi^{\prime}_{2})
- \sin\vartheta^{\prime}_{1} \sin\vartheta_2  \nonumber  \\
&& \times \cos(\varphi^{\prime}_{1} - \varphi_2) + \sin\vartheta^{\prime}_{1}
\sin\vartheta^{\prime}_{2} \cos(\varphi^{\prime}_{1} - \varphi^{\prime}_{2})|
= 2 \sqrt{2}.
\end{eqnarray}
Actually, having established that there exist infinitely many sets of directions satisfying the equality (17) for the singlet state, it is immediate to see that the same hols for the above equality (19) corresponding to the triplet state. Indeed, comparing the expressions in Eqs.\ (17) and (19), it follows at once that if the set of directions $\vec{e}_1(\vartheta_1,\varphi_1)$, $\vec{e}\,^{\prime}_{1}(\vartheta^{\prime}_{1},\varphi^{\prime}_{1})$, $\vec{e}_2(\vartheta_2,\varphi_2)$, and $\vec{e}\,^{\prime}_{2}(\vartheta^{\prime}_{2}, \varphi^{\prime}_{2})$ fulfill the equality (17), then the set of directions $\vec{e}_1(-\vartheta_1,\varphi_1)$, $\vec{e}\,^{\prime}_{1}(-\vartheta^{\prime}_{1}, \varphi^{\prime}_{1})$, $\vec{e}_2(\vartheta_2,\varphi_2)$, and $\vec{e}\,^{\prime}_{2}(\vartheta^{\prime}_{2}, \varphi^{\prime}_{2})$ do satisfy the equality (19) [alternatively, the set of directions $\vec{e}_1(\vartheta_1,\varphi_1)$, $\vec{e}\,^{\prime}_{1}(\vartheta^{\prime}_{1}, \varphi^{\prime}_{1})$, $\vec{e}_2(-\vartheta_2,\varphi_2)$, and $\vec{e}\,^{\prime}_{2}(-\vartheta^{\prime}_{2}, \varphi^{\prime}_{2})$ also will do].

A reasoning similar to that developed for the singlet and triplet states could equally be established for any other state of the form (5a) or (5b) for which $|c_1|=|c_2|=1/\sqrt{2}$ and $\vartheta_3 =\pi/2$, thereby showing that for every maximally entangled state there are infinitely many sets of directions giving the maximal violation of the CHSH inequality. As we have already said, such infinity of directions arises due to the {\it nonuniqueness\/} of the Schmidt decomposition of a maximally entangled state.

\section{Concluding remarks}

Lastly, a further remark is in order about the relationship between the conditions needed to maximally violate the CHSH inequality and those required by the three-particle state (2) to produce a direct (``all or nothing'') nonlocality contradiction \cite{GHZ,GHSZ}. As pointed out by KZ, it is remarkable that the same type of spin measurements involved in the two-particle case also forms the basis of the argument leading to the GHZ contradiction \cite{GHSZ}. This connection can be best appreciated when we generalize the CHSH inequality to a measurement scheme involving three spin-$\frac{1}{2}$ particles. The appropriate inequality is of the form \cite{Hardy}
\begin{equation}
|\langle B_{\text H} \rangle | \leq 2 \, ,
\end{equation}
where now the relevant Bell operator is
\begin{eqnarray}
B_{\text H} \equiv &[& \sigma(\vec{e}_1)\otimes \sigma(\vec{e}\,^{\prime}_{2}) +
\sigma(\vec{e}\,^{\prime}_{1})\otimes \sigma(\vec{e}_{2})]\otimes
\sigma(\vec{e}\,^{\prime}_{3})  \nonumber   \\
&& + \, [\sigma(\vec{e}\,^{\prime}_{1})\otimes \sigma(\vec{e}\,^{\prime}_{2}) -
\sigma(\vec{e}_1)\otimes \sigma(\vec{e}_{2})]\otimes
\sigma(\vec{e}_{3}) .
\end{eqnarray}
As in the two-particle case, it can be shown \cite{Cereceda} that in order for the three-particle state (2) to yield the largest violation of the inequality (20) it is necessary that $|c_1|=|c_2|=1/\sqrt{2}$. However, the largest eigenvalue of the Bell operator (21) is given by \cite{Cereceda}
\begin{eqnarray}
\lambda_l = && 2 \, (1+|\sin\theta_1 \sin\theta_2| + |\sin\theta_2 \sin\theta_3| \nonumber \\
&& + \, |\sin\theta_1 \sin\theta_3|)^{1/2} ,
\end{eqnarray}
so that a maximum violation occurs provided that $\vec{e}_i \perp \vec{e}\,^{\prime}_{i}$ for each $i=1,2,3$. The corresponding eigenvector will consist of an equally weighted superposition of the states $|\sigma_{\perp 1}\rangle|\sigma_{\perp 2}\rangle|\sigma_{\perp 3}\rangle$ and $|-\sigma_{\perp 1}\rangle|-\sigma_{\perp 2}\rangle|-\sigma_{\perp 3}\rangle$ with a relative phase between them. As before, $\theta_i$ is the angle between the vectors $\vec{e}_i$ and $\vec{e}\,^{\prime}_{i}$, and $|\sigma_{\perp i}\rangle$ ($|-\sigma_{\perp i}\rangle$) denotes the eigenvector [with eigenvalue $\sigma_{\perp i}=\pm 1$ ($-\sigma_{\perp i}=\mp 1$)] of the spin operator for particle $i$ along the direction perpendicular to both $\vec{e}_i$ and $\vec{e}\,^{\prime}_{i}$. Likewise, the relative signs of the eigenvalues $\sigma_{\perp 1}$, $\sigma_{\perp 2}$, and $\sigma_{\perp 3}$ entering the superposition will depend on the relative signs of $\sin\theta_1$, $\sin\theta_2$, and $\sin\theta_3$. [Note that the maximum amount of violation of inequality (20) predicted by quantum mechanics is by a factor of 2 instead of the factor $\sqrt{2}$ achieved in the CHSH inequality. This fact conforms to the existence of Bell type inequalities which yield a violation increasing exponentially with the number of particles \cite{Mermin1}.] The point we want to emphasize here is that, as shown by Hardy \cite{Hardy}, a maximum violation of inequality (20) always entails a contradiction of the GHZ type. A simple illustration of this statement is provided by the choice $\sigma(\vec{e}_i)\equiv \sigma(x_i)$ and $\sigma(\vec{e}\,^{\prime}_{i})\equiv \sigma(y_i)$, $i=1,2,3$, where $\sigma(x_i)$ [$\sigma(y_i)$] denotes the spin operator along the $x$ axis ($y$ axis) for particle $i$. For this choice of operators we find that $|\langle B_{\text H} \rangle| =4$ whenever the expectation value is evaluated for the state vector (1). As is well known, both the operators $\sigma(x_i)$ and $\sigma(y_i)$, and the state vector (1), form the basis of Mermin's exposition on the GHZ theorem \cite{Mermin2}.\footnote{%
Actually, the state employed by Mermin was $|\phi\rangle = 1/\sqrt{2}\,(|z+ \rangle_1 |z+ \rangle_2 |z+ \rangle_3 - |z- \rangle_1 |z- \rangle_2 |z- \rangle_3)$. For this state, and for the above choice of operators, we have $\langle B_{\text H} \rangle =4$, whereas for the state in Eq.\ (1) we have $\langle B_{\text H} \rangle =-4$. In any case, both $|\phi \rangle$ and the state vector (1) provide the maximum violation of inequality (20) in terms of the absolute value, and both states can lead to the GHZ contradiction.}
It is worth noting that the plane containing the measurement directions $\vec{e}_i$ and $\vec{e}\,^{\prime}_{i}$ ($i=1,2,3$) giving the largest violation of inequality (20) is fixed by the quantum state since, in contrast to the two-particle case, the triorthogonal decomposition in Eq.\ (2) is unique even if the coefficients $c_1$ and $c_2$ are equal \cite{Elby-Bub}.

We conclude by noting that the present treatment regarding the three-particle state (2) can be readily extended to $n$-particle states ($n \geq 3$) of the form
\begin{equation}
|\Psi\rangle = c_1 |z_1\rangle |z_2\rangle \cdots |z_n\rangle + 
c_2 |-z_1\rangle |-z_2\rangle \cdots |-z_n\rangle  .
\end{equation}
Indeed, by letting $|z_j\rangle^{\textstyle\ast}$ be
\begin{equation}
|z_j\rangle^{\textstyle\ast} = \cos\frac{\vartheta_j}{2} e^{-iz_{j}\varphi_{j}/2}
|z_j\rangle + z_j \sin\frac{\vartheta_j}{2} e^{iz_{j}\varphi_{j}/2} |-z_j\rangle ,
\end{equation}
with $j=N+1, N+2,\ldots,n,$ and $N=2,3,\ldots,\mbox{$n-1$},$ and projecting $|\Psi\rangle$ onto the direct product state $|z_{N+1}\rangle^{\textstyle\ast} |z_{N+2}\rangle^{\textstyle\ast} \cdots |z_{n}\rangle^{\textstyle\ast}$, one finds that the resulting $N$-particle state $|\Psi^{+}_{12 \cdots N} \rangle$,
\begin{equation}
p^{1/2}_{+} |\Psi^{+}_{12 \cdots N} \rangle = { }^{\textstyle\ast}\langle z_{N+1}|
{ }^{\textstyle\ast}\langle z_{N+2}| \cdots { }^{\textstyle\ast}\langle z_n ||\Psi\rangle ,
\end{equation}
is entangled for every choice of $|z_j\rangle^{\textstyle\ast}$ except for the case in which $\vartheta_j$ happens to be a multiple of $\pi$. In Eq.\ (25), $p_+$ is a normalization factor given by
\begin{equation}
p_+ = c^2_1 \prod^{n}_{j=N+1} \cos^2 \left( \frac{\vartheta_j}{2} \right)
+ c^2_2 \prod^{n}_{j=N+1} \sin^2 \left( \frac{\vartheta_j}{2} \right) ,
\end{equation}
while the state vector $|\Psi^{+}_{12 \cdots N} \rangle$ is found to be
\begin{eqnarray}
|\Psi^{+}_{12 \cdots N} \rangle & = & p^{-1/2}_{+} \left[ c_1 \cos\frac{\vartheta_{N+1}}
{2} \cos\frac{\vartheta_{N+2}}{2} \cdots \cos\frac{\vartheta_{n}}{2} \right. \nonumber  \\
&& \times |z_1\rangle |z_2\rangle \cdots |z_N\rangle + c_2 (z_{N+1})(z_{N+2}) \cdots
(z_n)  \nonumber  \\
&& \times \sin\frac{\vartheta_{N+1}}{2} \sin\frac{\vartheta_{N+2}}{2} \cdots \sin\frac{\vartheta_{n}}{2}   \nonumber  \\
&& \times e^{-i(z_{N+1}\varphi_{N+1}+z_{N+2}\varphi_{N+2}+ \cdots +
z_{n}\varphi_{n})}  \nonumber  \\
&& \times |-z_1\rangle |-z_2\rangle \cdots |-z_N\rangle \biggr] .
\end{eqnarray}
Clearly, expressions (26) and (27) reduce to Eqs.\ (6a) and (5a), respectively, when we take $N=2$ and $n=3$. Similarly, by taking the partial trace of $|\Psi\rangle\langle\Psi|$ over the states corresponding to particles $N+1, N+2,\ldots, n,$ one finds that the reduced density matrix associated with the remaining $N$-particle system is
\begin{equation}
\rho_{12 \cdots N} = c^2_1 |\phi^{+}_{12 \cdots N}\rangle\langle \phi^{+}_{12 \cdots N}|
+ c^2_2 |\phi^{-}_{12 \cdots N}\rangle\langle \phi^{-}_{12 \cdots N}| ,
\end{equation}
with $|\phi^{+}_{12 \cdots N}\rangle$ and $|\phi^{-}_{12 \cdots N}\rangle$ being $|z_1\rangle
|z_2\rangle \cdots |z_N\rangle$ and $|-z_1\rangle |-z_2\rangle \cdots |-z_N\rangle$, respectively. This implies that no violation of local realism can result from joint measurements performed on particles $1,2,\ldots,N$ alone if such measurements are made without any commitment to the results obtained for particles $N+1,N+2,\ldots,n$ (as a matter of fact, no actual measurements need to be performed on particles $N+1,N+2,\ldots,n$ if we are looking at the unconditional correlation function for particles $1,2,\ldots,N,$ since this latter is, by definition, fully independent of whatever measurements on particles $N+1,N+2,\ldots,n$). Indeed, it is not difficult to show that the quantum prediction for the unconditional correlation function of the results of spin measurements on particles $1,2,\ldots,N$ along respective directions $\vec{e}_1,\vec{e}_2,\ldots,\vec{e}_N$ is given by
\begin{eqnarray}
E_{12 \cdots N} \!\!\!&(&\!\!\!\vec{e}_1,\vec{e}_2,\ldots,\vec{e}_N)  \nonumber  \\
& = & \text{Tr}[\rho_{12\cdots N}
\sigma(\vec{e}_1)\otimes \sigma(\vec{e}_2)\otimes \cdots 
\otimes \sigma(\vec{e}_N)]   \nonumber  \\
& = & z_1 z_2 \cdots z_N \cos\vartheta_1 \cos\vartheta_2 \cdots \cos\vartheta_N ,
\end{eqnarray}
which generalizes Eq.\ (9). Thus we have proved that, for the class of $n$-orthogonal states in Eq.\ (23), the unconditional $N$-particle correlations are compatible with local realism for any $N=2,3,\ldots,n-1$.

However, by adequately generalizing the CHSH inequality to an $N$-measurement scheme \cite{Hardy}, one could equally prove \cite{Cereceda} that in order for the state (27) to yield the largest violation of the appropriate Bell inequality it is necessary that the $c$ numbers in front of $|z_1\rangle |z_2\rangle \cdots |z_N\rangle$ and $|-z_1\rangle |-z_2\rangle \cdots |-z_N\rangle$ have the same modulus.\footnote{%
Notice that, in contrast with the situation above concerning the density matrix $\rho_{12 \cdots N}$, any eventual description of particles $1,2,\ldots,N$ in terms of the pure state $|\Psi^{+}_{12\cdots N}\rangle$ is subject to the occurrence of the results $z_{N+1}, z_{N+2}, \ldots, z_{n}$ for spin measurements performed on particles $N+1,N+2,\ldots,n$ along the respective directions $\vec{e}_{N+1},\vec{e}_{N+2},\ldots,\vec{e}_n$ [see Eq.\ (25)]. Clearly the probability for such an event is $p_+$.}
Obviously, when applied to the state (23), this condition means that $|c_1|=|c_2|=1/\sqrt{2}$. This in turn implies that all observers $N+1,N+2,\ldots,n$ must perform spin measurements within the respective $x$-$y$ plane (i.e., $\vartheta_j =\pi/2$ for all $j$) if the remaining state $|\Psi^{+}_{12\cdots N}\rangle$ for the other $N$ particles is to violate maximally Bell's inequality. In any case, the amount of violation, if any, does depend on the value of measurement parameters $\vartheta_j$ and $\varphi_j$ attached to measuring apparatuses which can operate quite independent (e.g., in a spacelike separated region) from the corresponding apparatuses used to measure the spin of particles $1,2,\ldots,N,$ thus giving rise to a generic {\it entangled entanglement\/} of the kind contemplated by Krenn and Zeilinger \cite{KZ}. [The dependence of the entanglement on the $\vartheta_j$'s as well as the dependence on the relative phase of $|\Psi^{+}_{12\cdots N}\rangle$ on the $\varphi_j$'s are made explicit in Eq.\ (27).]

\end{multicols}
\pagebreak

\end{document}